\documentclass[usenatbib]{mn2e}

\usepackage{graphicx}
\usepackage[hypertex]{hyperref}

\def \eg {e.g.}
\def \ie {i.e.}
\def\spose#1{\hbox to 0pt{#1\hss}}
\def\ltsim{$\mathrel{\spose{\lower 3pt\hbox{$\sim$}}
        \raise 2.0pt\hbox{$<$}}$\thinspace}
\def\gtsim{$\mathrel{\spose{\lower 3pt\hbox{$\sim$}}
        \raise 2.0pt\hbox{$>$}}$\thinspace}
\newcommand{\thin }{\thinspace}

\newcommand{\lcdm}{$\Lambda$CDM}

\newcommand{\rtwentyfive}{${\rm R_{2500}}$}

\newcommand{\kms}{${\rm km\ s^{-1}}$}
\newcommand{\vrot}{$v_{eff}$}
\newcommand{\veff}{\vrot}
\newcommand{\src }{NGC\thin 4649}
\newcommand{\zfe }{${\rm Z_{Fe}}$}

\newcommand{\chandra }{{\em Chandra}}

\newcommand{\astroh}{\href{http://astro-h.isas.jaxa.jp/}{{\em Astro-H}}}

\newcommand{\xspec }{{\em Xspec}}

\newcommand{\xmm }{{\em XMM}}

\newcommand{\rosat }{{\em Rosat}}

\newcommand{\fnonthermal}{$f_{nth}$}

\defcitealias{humphrey06a}{H06}
\defcitealias{humphrey08a}{H08}
\defcitealias{humphrey11a}{H11}
\defcitealias{brighenti09a}{B09}

\title[Reconciling stellar dynamics and X-rays]{Reconciling stellar dynamical and hydrostatic X-ray mass measurements of an elliptical galaxy with gas rotation, turbulence and magnetic fields}
\author[P.~J. Humphrey et al.]{\parbox{\textwidth}{Philip J. Humphrey$^1$,  David A. Buote$^1$, Fabrizio Brighenti$^{2,3}$, Karl Gebhardt$^4$ and William G. Mathews$^3$} \vspace{0.4cm}\\
\parbox{\textwidth}{$^1$ Department of Physics and Astronomy, University of California, Irvine, 4129 Frederick Reines Hall, Irvine, CA 92697-4575\\$^2$ Dipartimento di Astronomia, Universit\`{a} di Bologna, Via Ranzani 1, Bologna 40127, Italy\\$^3$ University of California Observatories, Lick Observatory, University of California at Santa Cruz, Santa Cruz, CA 95064\\$^4$ Astronomy Department, University of Texas, Austin, TX 78712}}
\begin{document}
\maketitle
\begin{abstract}
Recent hydrostatic X-ray studies of the hot interstellar medium (ISM)
in early-type galaxies  underestimate the gravitating mass as compared to 
stellar dynamics, implying modest, but significant
deviations from exact hydrostatic equilibrium. We present a method for 
combining X-ray measurements and stellar 
dynamical constraints in the context of Bayesian statistics that allows 
the radial distribution of the implied nonthermal pressure or bulk
motions in the hot ISM to be constrained. 
We demonstrate the accuracy of the method 
with hydrodynamical simulations tailored to produce a 
realistic galaxy model.
{Applying the method to the nearby elliptical galaxy \src, we find 
a significant but subdominant nonthermal pressure fraction ($0.27\pm0.06$)
in the central (\ltsim 5~kpc) part of the galaxy, 
similar to the level of deviations from hydrostatic equilibrium expected in 
galaxy clusters. Plausible sources of systematic error, if important, 
may reduce this fraction.}
Our results imply 
$\sim$360\kms\ random turbulence or a magnetic field
$B=(39\pm6)(n_e/0.1\ cm^{-3})^{0.59\pm0.09}\mu$G, whereas
gas rotation alone is unlikely to explain the detailed
nonthermal profile. Future observations with \astroh\ will allow 
turbulence or gas rotation at this  level to be detected.
\end{abstract}
%\keywords{Xrays: galaxies--- galaxies: elliptical and lenticular, cD--- galaxies: ISM--- methods: data analysis--- galaxies: individual (NGC 4649)}
%DONE
\section{Introduction}
The distribution of mass in early-type galaxies is a fundamental yardstick for understanding
their formation and evolution. Our current \lcdm\ cosmological paradigm predicts
ubiquitous, massive dark matter halos, the distribution of matter within which 
correlates with the mass of the system \citep[\eg][]{navarro97,bullock01a,maccio08a}.
Within the optical radius of the galaxy, the relative contributions to the gravitational 
potential of the dark and 
luminous matter may explain the tilt of the fundamental
plane \citep[\eg][]{cappellari06a,bolton07a,humphrey10a}, and provide insights into the 
poorly understood processes by which they interact gravitationally, in particular the interplay 
between adiabatic contraction and dynamical friction 
\citep[\eg][]{blumenthal86a,gnedin04a,elzant04a,humphrey06a,dutton07a,gnedin07a,abadi10a,napolitano10a,maccio12a}.
The stellar mass-to-light (M/L) ratio can provide important constraints on the 
history of star formation and the shape, and universality, of the stellar initial mass function (IMF)
\citep[\eg][]{gerhard01a,gnedin07a,humphrey09d,treu10a,cappellari12a}. Furthermore, at the smallest scales,
the tight relations between the central supermassive 
black hole (SMBH) mass and the global properties of the host galaxy
imply strong evolutionary symbiosis \citep[\eg][for a review]{kormendy95a,gebhardt00a,merritt01a,gultekin09a,ferrarese05a}.

A number of techniques are available to study the mass distributions in early-type galaxies,
each having its own distinct advantages and disadvantages. For example, 
gravitational lensing studies directly probe the projected mass distribution, but alone they do not
generally produce detailed mass profiles for individual giant elliptical galaxies, although they 
can provide important constraints on the profile averaged over multiple systems
\citep[\eg][]{gavazzi07a}, or when combined with stellar dynamics measurements 
\citep[\eg][]{koopmans09a,barnabe11a}. By combining information from 
multiple dynamical tracers, sophisticated (axisymmetric) orbit-based
(stellar) dynamical models are now able to probe from within the sphere of influence of the 
SMBH out to tens of kpc \citep[\eg][]{romanowsky03a,gebhardt09a}, but models
self-consistently incorporating dark matter halos can be computationally costly to evaluate
\citep[\eg][]{shen10a}, and care must be taken to minimize systematic uncertainties 
associated with their implementation \citep[\eg][]{valluri04a,shen10a,das11a,long12a}
and the (generally) unknown inclination of the system \citep{gavazzi05a,thomas07b}.

Hydrostatic X-ray techniques,
on the other hand, are appealing in part due to their
computational simplicity, given the isotropy of the gas pressure tensor
and the fact that the spherical approximation \citep[which is generally assumed, although
not exclusively; \eg][]{buote94,buote96a,buote98d,buote02b,statler02a} 
typically introduces only a small bias
\citep[][and references therein]{buote11c}, particularly
if the spherically averaged mass profile is close to a singular isothermal sphere 
\citep{buote11b,churazov08a}, which is generally true in early-type galaxies 
\citep[\eg][]{koopmans09a,humphrey10a,churazov10a}. These 
methods are especially powerful since current instrumentation allows measurements from
close to the sphere of influence 
of the most massive nearby black holes out to \gtsim\rtwentyfive\ for some 
nearby galaxies \citep[][for a review]{humphrey08a,humphrey09d,humphrey11a,humphrey12b,wong11a,buote11a},
and the same approach can be applied self-consistently from $\sim$Milky Way-mass galaxies to
massive clusters. 

The accuracy of hydrostatic methods is, naturally, 
contingent on the extent to which  equilibrium gas motions 
reflect the thermal gas pressure rather than turbulent or 
streaming motions. 
To ensure systems are not grossly far from equilibrium and do not 
exhibit dynamically dominant, supersonic gas motions, excluding objects
with highly asymmetric X-ray isophotes is an important first step
\citep{buote11a}. Nevertheless, subsonic gas motions (either bulk or 
turbulent), or nonthermal support from magnetic fields or cosmic rays
could still be important \citep[\eg][]{churazov08a}. 
Cosmic ray pressure is likely to be 
most important in the vicinity of cavities inflated by an AGN jet
\citep[\eg][]{mathews08b}. Magnetic fields weaker than a few 
$\mu$G in the centre of the galaxy should not be dynamically 
important \citep{buote11a}, but there is considerable uncertaintiy
on the actual magnetic field strength, with
current observational constraints from
galaxies with embedded radio jets ranging from $\sim$1$\mu$G to a 
few tens of $\mu$G \citep{vallee11a}. 
%DONE

Interesting constraints on 
gas dynamics are few \citep{buote11a}. The upper limits on X-ray spectral line broadening 
\citep[$\sim$400--500\kms, from \xmm\ RGS observations:][]{sanders11a},
and velocity gradients \citep[$\sim$1000--2000\kms\ from the spatial
variation of the X-ray line centroids in clusters:][]{ota07a,sato08b,sugawara09a,tamura11a}
exceed the typical circular velocity for a giant elliptical galaxy.
Indirect gas dynamical constraints in the very centres of early-type galaxies can be 
obtained from  resonance scattering (or 
its absence) in strong emission lines \citep[][for a review]{churazov10b}. 
Current measurements suggest microturbulent velocities ranging from 
$\sim$300--700\kms\ and 100--500\kms\ respectively in 
two giant elliptical galaxies \citep[albeit sensitive to the fitted 
plasma code;][]{deplaa12a}, to \ltsim 100\kms\ in the centre
of the elliptical galaxy NGC\thin 4636 (\citealt{xu02a}; \citealt{werner09a},
who also found evidence of little turbulence in two other galaxies). 
The interpretation of these results,
however, may be complicated by the scale and anisotropy of the 
turbulence \citep{zhuravleva11a}. Dynamically important gas rotation, such
as expected in subsonically inflowing gas due to angular momentum conservation,
should result in a flattening of the X-ray isophotes parallel to the rotation axis.
This allowed \citet[][hereafter \citetalias{brighenti09a}]{brighenti09a} to infer $\sim$200--300\kms\ gas motions
in the central $\sim$kpc of the elliptical galaxy NGC\thin 4649, and 
led them to suggest that inflowing, rotating gas may be common in the very
central parts of nearby elliptical galaxies.

A direct assessment of the accuracy of hydrostatic methods
in an elliptical galaxy can be made by comparing the mass inferred
from the X-rays to that obtained by  independent techniques. 
Orbit-based stellar dynamical models are the alternative method of
choice for nearby systems, as they provide
high-quality mass profiles that overlap well in
radial scale with X-ray work. To date such comparisons have been 
made reliably in only a handful of cases. \citet{gebhardt09a} found 
inconsistency between their axisymmetric dynamical models for M\thin 87
and the gravitational potential inferred from X-rays by \citet{churazov08a}.
\citet{das10a} and \citet{murphy11a} quantified this discrepancy 
further, finding that the X-rays underestimate the optically inferred 
mass by $\sim$50\%\ within the central $\sim$4~kpc, while the data
generally agreed at larger scales. M\thin 87 exhibits, however, 
substantial disturbances to its 
X-ray morphology \citep[\eg][]{bohringer94a,forman07a,million10a}, 
which may introduce systematic 
uncertainties into a single-phase spherical hydrostatic mass analysis.
\citet{shen10a} found that the X-ray inferred mass that was measured by 
\citet{humphrey08a} for the elliptical galaxy NGC\thin 4649 
was $\sim$40\%\ lower than their dynamical models implied
inside $\sim$8~kpc, despite its very smooth, relaxed
X-ray isophotes. A similar result was obtained by \citet{das10a},
who also compared their X-ray determined mass profile for NGC4472 to (unpublished)
orbit-based models, finding it to be $\sim$30\%\
lower inside $\sim$10~kpc but $\sim$20\%\ {\em higher} further out. 
This dichotomy may, in part, reflect X-ray disturbances in the system at the largest 
scales \citep{irwin96a,biller04a}. \citet{rusli11a} measured the mass 
distribution in the central regions of the lenticular galaxy 
NGC\thin 1332 with axisymmetric orbit-based models (excluding a 
dark matter halo), and finding the 
X-ray inferred profile obtained by \citet{humphrey09d} to be $\sim$50\%\ lower.

A number of other studies are less definitive due to limitations in the optical 
or X-ray analysis. For example, \citet{norris12a} derived the mass of NGC\thin 3923 from 
axisymmetric orbit-based models, including a dark matter halo,
finding an offset from the X-ray mass profile of \citet{fukazawa06a}. Although
corrections were applied to account for unresolved sources when computing 
the temperature, \citeauthor{fukazawa06a} did not account for this contamination
when deriving the density profile. An independent \rosat\ study by 
\citet{buote98d}, considering non-spherical mass models for this galaxy,
was actually in better agreement with the dynamical mass. Several authors have compared 
X-ray inferred masses with simpler dynamical models that may be less reliable
\citep[\eg][]{mathews03b,ciotti04a,churazov08a,churazov10a,romanowsky09a,johnson09a,das10a},
or have used X-ray mass profiles that are likely unreliable due to large morphological
disturbances (\eg\ NGC\thin 4636: \citealt{johnson09a}; NGC\thin 5846: 
\citealt{churazov10a,das10a}), unresolved point source contamination \citep{pellegrini06a},
or systematic uncertainties associated with the model parameterizations 
\citep{romanowsky09a}.

Based on these studies, a consistent picture is beginning to 
emerge of X-ray hydrostatic methods underestimating the mass inferred from
stellar dynamical techniques by {\em on average} 
$\sim$30--50\%\ in the central part of the galaxy. Understanding this
discrepancy is essential not only in order to correct hydrostatic
mass for this effect, but also to gain insight into the behaviour
and fate of the gas at the centre of a galaxy-scale cooling flow
\citep{mathews03a}. Obtaining detailed radial profiles of the measured nonthermal
pressure or bulk gas motions is a critical next step. 
While, in essence, this can be achieved
by differencing the mass profiles inferred by both techniques, most recent 
studies have only compared them approximately
\citep[\eg][]{shen10a,das10a,das11a,rusli11a,murphy11a,norris12a}\footnote {We note that if one compares the difference between two radii in the {\em potential} (rather than the enclosed mass) that is obtained with each method, as advocated by 
\citet{churazov08a}, the average nonthermal
pressure in that range can also be derived. The interpretation is 
complicated, however, by the lack of statistical independence between the potential
``data-points'' generated in this way. Nevertheless, \citeauthor{churazov08a}
used this approach to infer the {\em global average} level of support in 
two objects.}

Given that the level of discrepancy is \ltsim 50\%, 
it is crucial to control
tightly the systematic uncertainties in both the X-ray and stellar dynamical measurements.
From an X-ray perspective, there are a number of ways to infer the gravitating mass
from X-ray data, as reviewed in 
\citet{buote11a}. For a reliable measurement, however, 
any fitted models must have sufficient flexibility to capture the full range of physical 
gas temperature and density profiles. 
The traditional ``smoothed inversion'' technique, as widely employed in 
studying galaxy clusters, involves fitting parameterized models directly to the 
density and temperature distributions but, due to their diverse shapes in 
elliptical galaxies, there are in general no well-defined ``universal'' parameterizations
that can be employed, introducing potentially large systematic uncertainties 
in this regime \citep[\eg][]{humphrey09d,buote11a}. 
\citet{das10a} proposed a minimally parametric ``smoothed inversion'' technique 
that does not {\em a priori} restrict the form of the density and temperature profiles,
{but it does employ a certain level of smoothing. We discuss the performance of this
method in the Appendix.}
In practice, we have found an entropy-based ``forward fitting'' approach to be an attractive
means for computing the mass profile reliably, as it rigorously enforces both a physical mass
distribution and Schwarzschild's stability criterion for the ISM 
\citep{humphrey08a,humphrey09d,humphrey11a}. The drawback of this technique for 
computing the nonthermal pressure profile is that it assumes the mass inferred by 
hydrostatic techniques can be well-parameterized by the same model that can fit 
the true mass distribution, which need not be true in the case of a nonthermal pressure
gradient.

In this paper, we develop a simple extension of the entropy-based, forward fitting
technique that incorporates non-hydrostatic effects. This method allows 
the radial distribution of the nonthermal pressure or gas motions to be 
constrained directly from a joint, Bayesian analysis of the 
X-ray and stellar dynamics data. We illustrate the performance of the method
both with simulated galaxies, and 
with a real system (\src), and interpret the results in terms of physical models
for the gas flow, turbulence and possible magnetic fields.
In what follows, all error bars correspond to 1-$\sigma$ uncertainty.
%DONE 

\section{NGC\thin 4649} \label{sect_n4649}
\begin{figure}
\centering
\includegraphics[height=3.5in,width=3in,angle=270]{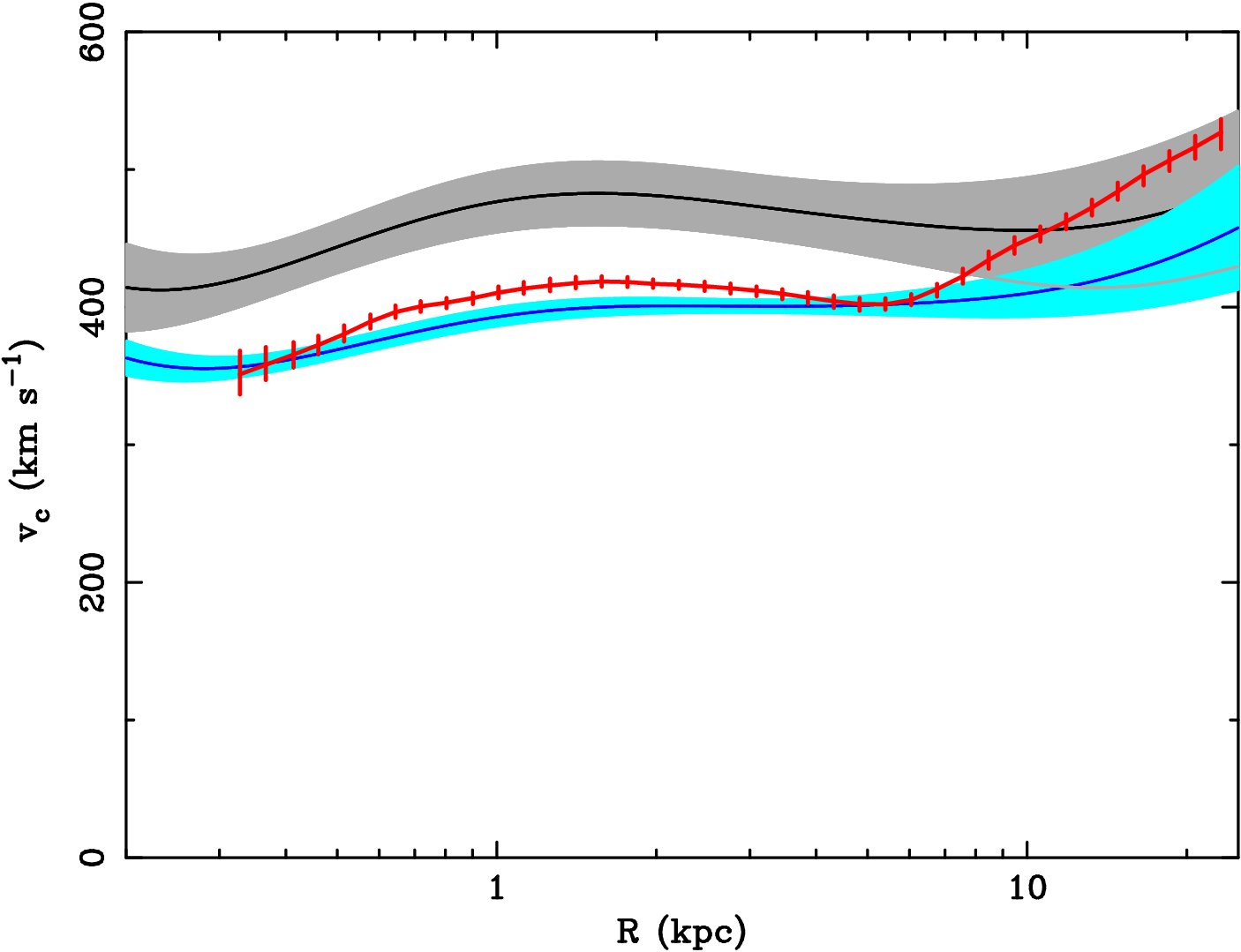}
\caption{Comparison of the cicular velocity profile of \src\ derived from the 
stellar dynamical analysis of \citet[][grey region]{shen10a}, and our 
purely hydrostatic analysis of the hot ISM (blue shaded region). We also show (red) the 
best-fitting model and 1-$\sigma$ confidence region obtained by  \citet{das10a}
from their hydrostatic X-ray analysis. The modest discrepancies between the 
X-ray measurements likely arise from different modelling assumptions 
in the \citeauthor{das10a}
analysis (see the Appendix), while the higher optical measurement can 
be interpreted as arising from modest nonthermal pressure.\label{fig_vc}}
\end{figure}
To test our method, and provide important constraints on a well-studied
system, we focused our analysis on \src. 
This is a nearby (15.7~Mpc)\footnote{We adopt a distance of 15.7~Mpc for 
consistency with the dynamical models of \citet{shen10a}}, giant elliptical 
galaxy with a luminous X-ray halo and a remarkably round, relaxed X-ray
morphology, at least within $\sim$20~kpc 
\citep{trinchieri97b,randall03,humphrey08a}. Low significance surface brightness
asymmetries were reported in a shallow \chandra\ observation by \citet{randall03},
which they, in part, speculated may arise due to convective flows. This hypothesis
is inconsistent, however, with the monotonically rising entropy profile (implying global 
stability against convection). In any case,
the features were not confirmed in deeper \chandra\ observations \citep{humphrey08a}. 
\citet{shurkin07a} reported the detection of small cavities associated with the
weak radio jet but these were also not confirmed in our deeper data.
Instead the gas was found to be very symmetrical and relaxed-looking 
\citep{humphrey08a}. The X-ray isophotes do exhibit central flattening, however,
which suggests that gas rotation may be important in the central $\sim$1~kpc
\citepalias{brighenti09a}.

Given its proximity and high surface brightness, we were able to measure the gas properties
from close to the sphere of influence of the central black hole ($\sim$1\arcsec) out
to $\sim$25~kpc, allowing, for the first time, the mass of {\em any} SMBH to be inferred
directly from a hydrostatic X-ray model \citep{humphrey08a}.
State of the art, orbit-based, axisymmetric stellar dynamical models, which spanned a similar
radial range, were constructed by \citet{shen10a}, assuming an edge-on geometry
(but verifying that the results were not very sensitive to the adopted inclination).
Although the inferred black hole mass was in statistical agreement with the X-ray 
measurement, they found a modest discrepancy with 
the hydrostatic mass profile within the central $\sim$8~kpc. A similar discrepancy was
noted by \citet{das10a} between the dynamical mass of 
\citeauthor{shen10a} and their own X-ray mass model.

In Fig~\ref{fig_vc}  we show the circular velocity ($v_c=\sqrt{G M/r}$, where G 
is the Universal gravitational constant and M is the total mass enclosed within 
radius r) profile for \src, based on the X-ray data \citep{humphrey08a} and the 
dynamical analysis of \citet{shen10a}\footnote{This Bayesian realization of the $v_c$ 
profile  was derived by linearly interpolating over the $\chi^2$-topology 
obtained by \citet{shen10a} from their stellar dynamical analysis.},
illustrating the offset between the inferred mass distributions. We also overlay
the $v_c$ profile of \citet{das10a}, which overall agrees well with our measurement
over the range $\sim$3--9~kpc, but is marginally higher outside this range. We 
ascribe these differences to particular choices 
made in their analysis (see the Appendix 
for a more detailed discussion). The generally good agreement between the
two X-ray measurements indicates that the level of discrepancy with the 
optical results is
not sensitive to the details of the X-ray modelling although, as discussed in the Appendix,
the error bars may be.

In the present study, we interpret 
the offset between the X-ray and dynamical measurements as wholly arising due to 
deviations from hydrostatic equilibrium. Naturally, this is contingent upon the 
optical result being an unbiased measurement of the true gravitational potential. 
{For the state of the art models used by \citet{shen10a}, which have
previously shown to perform well in recovering the mass of simulated galaxies, 
we do not expect there to be significant inherent systematic uncertainties,
other than a dependence on the viewing angle \citep{thomas07b}. We therefore
assume that the gravitational potential inferred by \citet{shen10a} is 
accurate. We note that, recently, \citet{das11a} constructed  ``made to measure'' 
particle-based models for the system (using the same kinematics data in the central
region), finding larger statistical errors than \citeauthor{shen10a}, and 
thus formal consistency with the X-ray profile. However, investigating the origin
of such systematic differences is beyond the scope of the present paper.}

In order to constrain the nonthermal pressure profile, we carried out a joint
analysis of the optical and X-ray data, as described in
\S~\ref{sect_method}. For the X-ray data, we used the 
published gas density and temperature profiles from \citet{humphrey08a}.
For the optical constraints, we used the posterior probability distribution
obtained by \citet{shen10a} 
and linearly interpolated over their grid of $\chi^2$ values (and 
assumed flat priors for each parameter).

\section{Measuring nonthermal pressure}
\subsection{Preliminaries}
If the ISM is in an equilibrium state and the magnetic fields are
tangled, the total gravitating mass, M, enclosed within a surface ${\bf S}$ is 
given by \citep{fang09a}:
\begin{equation}
M = \frac{1}{4\pi G}\int_{\bf S} \left[ -\frac{1}{\rho_g} \nabla P - \left({\bf v}.\nabla\right){\bf v}\right] . {\bf dS}
\end{equation}
where $G$ is the universal gravitational constant, $\rho_g$ is the gas density,
$P$ is the total pressure (including thermal gas pressure, and pressure from 
tangled magnetic fields,
cosmic rays and turbulence), ${\bf v}$ is the (time averaged) gas velocity field 
and ${\bf dS}$ a surface element. Assuming ${\bf S}$ is a spherical surface of 
radius r, this becomes:
\begin{equation}
\frac{v_c^2}{r} = \frac{1}{4 \pi} \int_{4\pi } -\frac{1}{\rho_g} \frac{dP}{dr}  d\Omega - 
\frac{1}{4\pi} \int_{4\pi} {\bf \hat r} . \left({\bf v}.\nabla \right) {\bf v}\  d\Omega
\label{eqn_vc}
\end{equation}
where $v_c$ is the circular velocity, $\Omega$ is the solid angle, and 
${\bf \hat r}$ is the unit vector in the radial direction.
{For convenience, we define $<\rho_g>$ such that 
\begin{eqnarray}
\frac{1}{<\rho_g>}\frac{d<P>}{dr} & \equiv &   \frac{1}{4 \pi} \int_{4\pi } \frac{1}{\rho_g} \frac{dP}{dr}  d\Omega \label{eqn_p_average}
\end{eqnarray}
where $<P>$ is the spherically averaged pressure. We also define an ``effective'' gas
velocity, $v_{eff}$, such that 
\begin{eqnarray}
\frac{v_{eff}^2}{r} & \equiv & - \frac{1}{4\pi} \int_{4\pi} {\bf \hat r} . \left({\bf v}.\nabla \right) {\bf v}\ d\Omega \label{eqn_veff}
\end{eqnarray}
To make further progress, we assume that the density and pressure are approximately 
stratified on concentric spheres, which is generally a good approximation
\citep[][and references therein]{buote11b,buote11c}. In that case, 
we can drop the $<>$ averaging notation in Eqn~\ref{eqn_p_average}.}
If the only 
dynamically important gas motions are rotation about a common axis 
(\ie\ ${\bf v}=v_\phi {\bf e_\phi}$ where ${\bf e_\phi}$ is the unit vector in the
$\phi$ direction in spherical coordinates), 
$v_{eff}^2=\int_{4\pi} v_\phi^2 d\Omega / 4\pi$; hence $v_{eff}$ would be a spherically
averaged gas rotation velocity.

We define $P\equiv P_g/(1-f_{nth})$, where $P_g$ is the thermal gas pressure 
($P_g=\rho_g kT/(\mu m_H)$; k is Boltzmann's constant, T is the gas 
temperature, $\mu \simeq 0.62$ is the mean molecular weight, 
and $m_H$ is the weight of hydrogren), and \fnonthermal\ is the nonthermal pressure
fraction. We also define $S=\rho_g^{-2/3} kT/(\mu m_H)$, which is proportional to the 
traditional definition of the entropy proxy, $K= n_e^{-2/3} kT$, where
$n_e$ is the electron number density. Combining
Eqns~\ref{eqn_vc}--\ref{eqn_veff} and rearranging, we obtain
\begin{equation}
r^2 \left(\frac{S}{P_g}\right)^\frac{3}{5} \frac{d}{dr} \left( \frac{P_g}{1-f_{nth}} \right) = - r\left[ v_c^2-v_{eff}^2\right] \label{eqn_he2}
\end{equation}
Writing $v_c^2=v_{c,g}^2+v_{c,ng}^2$, where $v_{c,g}$ is the circular velocity
due to the gas mass, and $v_{c,ng}$ is due to the non-gas mass, differentiating
and folding in the gas mass continuity equation,
\begin{equation}
\frac{d}{dr} \left( r v_{c,g}^2\right) = 4 \pi G r^2 \left( \frac{P_g}{S} \right)^\frac{3}{5}
\end{equation}
we obtain:
\begin{eqnarray}
\frac{d}{dr}\left[ r^2 \left(\frac{S}{P_g}\right)^\frac{3}{5} \frac{d}{dr} \left( \frac{P_g}{1-f_{nth}} \right) \right] & = &  -\frac{d}{dr} r\left[ v_{c,ng}^2-v_{eff}^2\right] \nonumber \\ 
& & -  4 \pi G r^2 \left( \frac{P_g}{S} \right)^\frac{3}{5} \label{eqn_he}
\end{eqnarray}
%DONE 
\subsection{Method} \label{sect_method}
Adopting parameterized models for $S$, $v_{c,ng}$, \fnonthermal\ and $v_{eff}$, Eqn~\ref{eqn_he}
can be integrated numerically, to solve for P using a fourth-order Runge-Kutta method. 
We start the integration at some arbitrary, small radius $r_0$. Two
boundary conditions are needed; we set the thermal gas pressure at $r_0$ to a value
to be determined by our fit, and assume that the gas mass is zero within $r_0$.
Since the model profiles of $P_g$ and $S$ uniquely determine the gas density and 
temperature, the observed
temperature and density profiles can be fitted  by adjusting the various model parameters.
As the temperature and density are generally inferred by spherical
deprojection of spectra accumulated in a series of concentric bins, 
to ensure a reliable comparison
we integrated appropriately weighted functions of pressure and temperature
to compute the predicted contribution to the emission
measure (and hence mean density) from gas in the corresponding
shell, as well as an emission-weighted temperature. We compared these averaged
quantities to our measured temperature and density data points,
as discussed in \citet{humphrey08a} and \citet{humphrey09d}.
See Appendix B of \citep{gastaldello07a} for a detailed discussion
of incorporating the plasma emissivity into the gas modelling.

To parameterize $v_{c,ng}$, we model the gravitating mass as arising from a 
central black hole, a stellar mass component (assuming mass follows light),
plus a dark matter halo, which can either be a \citet[][hereafter NFW]{navarro97}
profile, or a ``cored logarithmic'' potential \citep{binney08a}.
The black hole 
mass, stellar M/L ratio, and normalization and characteristic scale
of the dark matter component are 
adjustable fit parameters. To parameterize $S$, we adopt a multiply broken
powerlaw, plus a constant term. This model has sufficient flexiblity 
to capture the overall shapes of the entropy profiles in galaxy 
groups and clusters \citep[\eg][]{gastaldello07b,sun09a,cavagnolo09a,humphrey12a}.
Since {\em a priori} there is no obvious parameterized form for 
\fnonthermal\ or $v_{eff}$, we parameterize them with a cubic spline function.
While the number of spline knots is arbitrary, we found the model to be 
generally well-behaved if there are roughly half as many knots as there are density
(or temperature) bins, and they are placed approximately logarithmically in radius. 
Since \fnonthermal\ only appears inside the derivative in
Eqn~\ref{eqn_he}, a boundary condition on it is required. We therefore
set \fnonthermal$=0$ at large radii.

In the hydrostatic approximation,
$f\equiv0$ and $v_{eff}\equiv 0$ and so the two unknowns ($S$ and $v_c$) can be 
constrained completely by the two observables ($\rho_g$ and $kT$).
By exploring parameter space with a Bayesian code 
(specifically, we use 
version 2.7 of the MultiNest code\footnote{{http://www.mrao.cam.ac.uk/software/multinest/}} \citep{feroz09a}), we can also fold in stellar dynamical constraints on 
$v_c$ by using the posterior probability distribution from the dynamical
analysis as priors on some of the parameters. In this case, we can relax the hydrostatic
approximation and additionally measure the profile of {\em either} 
\fnonthermal\ or $v_{eff}$. These two quantities are highly degenerate and so, without 
additional constraints, they cannot be disentangled uniquely. We therefore
adopt the pragmatic approach of fitting each profile in turn, while setting the
other term to zero. 
%DONE

\subsection{Tests} \label{sect_simulations}
\begin{figure}
\centering
\includegraphics[height=3.5in,width=3in,angle=270]{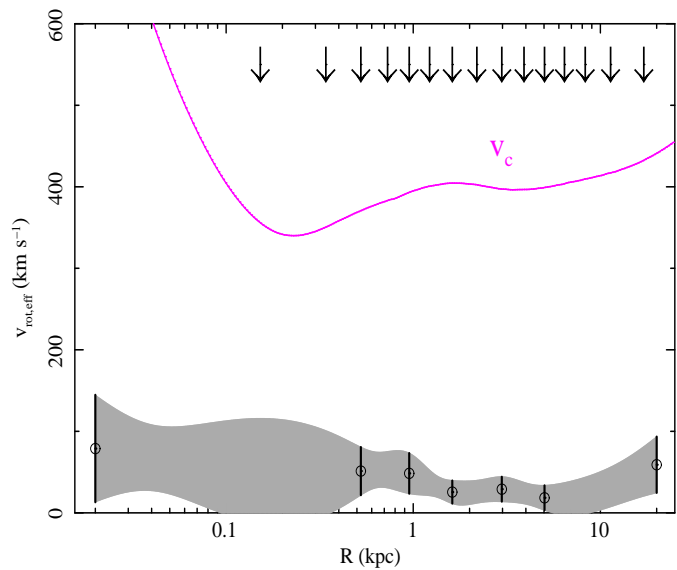}
\caption{Recovered \veff\ for the hydrostatic simulation. The 
data-points correspond to the spline knots fitted to the simulated data,
and the grey region indicates the 1-$\sigma$ confidence region of the model
interpolated between them.  
%The red shaded region is the distribution of \veff\
%recovered if the central data-bins are excluded. 
The magenta line is the circular velocity
($v_c$; {set to match the X-ray mass inferred by \citealt{humphrey08a}}), and the arrows indicate the approximate centre of each
annulus used in the spectral analysis\label{fig_faked_nonthermal}. The true 
\veff\ distribution (\veff$=0$) is consistent with the best fitting
model, although there is a slight ($\sim$30\kms) bias, which is not 
dynamically important.}
\end{figure}
\begin{figure*}
\centering
\includegraphics[width=7in]{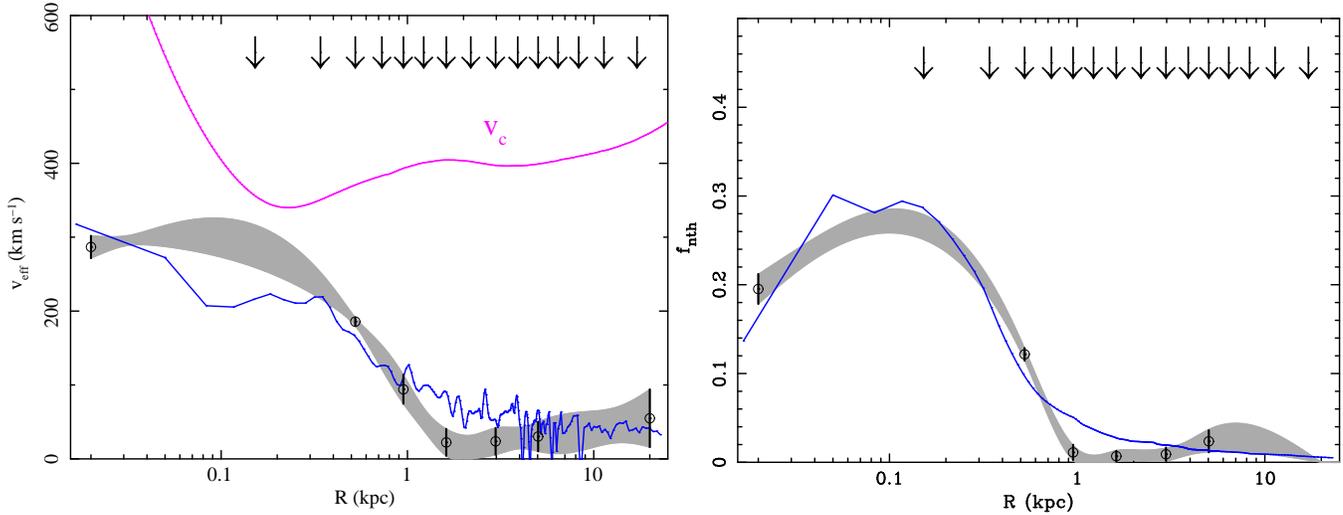}
\caption{ {\em Left:} Marginalized \veff\ from fitting the B09 simulation.
The data-points indicate the spline knots fitted to the simulation, and the grey
region is the 1-$\sigma$ confidence region of the model interpolated between them.
The blue line indicate the true profile of 
\veff. The magenta line is the circular velocity {\citep[matching the mass inferred by][]{humphrey08a}},
and the arrows indicate the approximate centre of each annulus used in the 
spectral analysis. Overall, we find good agreement between the true distribution and 
the model. {\em Right:} The same nonthermal support profile parameterized as a 
nonthermal pressure fraction (\fnonthermal; grey shaded region and data-points), 
shown with the true profile (blue) and (arrows)
the approximate centres of each radial bin. The overall fit is also 
fairly good . {We note that the modest discrepancy between the true and 
inferred $v_{eff}$ profiles around $\sim$0.1--0.2~kpc occurs between two spline
knots and is mostly confined to the central radial bin, so we caution against
over-interpretation.}
%In this case, the fit is good 
%except in the central bin, where the strong peak around $\sim$0.1~kpc cannot
%be captured by the adopted parameterization. This scale is 
%unresolved by the \chandra\ data, so we consider the overall
%agreement good. 
\label{fig_simulated_nonthermal}}
\end{figure*}
To test the ability of our method to recover the nonthermal pressure profile
in early-type galaxies, we used it to investigate two simulated datasets, 
each having different
amounts of nonthermal support. Since our method involves directly fitting the 
X-ray data, and since we assume that the
stellar dynamical constraints provide an unbiased estimate of the true
gravitational potential, we considered the limit of perfect optical constraints,
but a typical X-ray exposure (100~ks).
The first of these datasets was derived assuming 
spherical symmetry and perfect hydrostatic equilibrium
 (\ie\ \veff$\equiv$\fnonthermal$\equiv 0$; hereafter the ``hydrostatic simulation''). 
The second simulated dataset
was based on the axisymmetric, hydrodynamical simulation 
performed by \citetalias{brighenti09a}, in which there is 
turbulence and bulk gas motion.
In both cases, the simulations were tailored to 
match approximately the properties of NGC\thin 4649, as observed with \chandra. 

For each 
simulated dataset, we started with the temperature, density and abundance of the gas
as a function of position in an axisymmetric (R,z) grid, and created an artificial
\chandra\ events file using a Monte Carlo method to generate photons. 
We assumed the source is at 15.7~Mpc, and the observation was 100~ks in duration.
At each grid position, a random number of photons were generated, 
assuming an APEC plasma with the appropriate temperature and gas abundance, and 
using representative \chandra\ responses for close to the aimpoint. Each photon was 
randomly assigned a PHA (energy) bin and a spatial position within the bin, 
before being projected onto the 
sky. We processed the simulated events file similarly to the real \chandra\ data of 
\src\ \citep{humphrey08a}, which involves spherical deprojection with the 
{\em projct} model in \xspec\ {\em vers 11}, and fitting an APEC plasma model 
with variable abundances, and produces the temperature, density and abundance profiles.
Where appropriate, we tied the abundance between adjacent annuli, to improve
constraints. To simplify the analysis, we 
filtered out photons originating from outside the outermost shell used in our 
deprojection. 

We fitted the resulting temperature and density profiles simultaneously with our 
modified hydrostatic model, assuming flat priors on all fit parameters. 
In these tests, we aimed to determine how accurately
the nonthermal pressure profile can be recovered in the limit of perfect
dynamical data, so we
fixed the mass model parameters to match the true distribution used in the simulations.
For the hydrostatic simulation, we show the recovered distribution of \veff\ 
in Fig~\ref{fig_faked_nonthermal}, demonstrating that the true distribution
(\veff$=0$) was reasonably well recovered. There was a slight 
($\sim 30 km\ s^{-1}$) bias in the best-fitting velocity, 
which may reflect slight uncertainties in the deprojection procedure, 
but such a small velocity would be dynamically unimportant and so has 
minimal effect on our conclusions.

In Fig~\ref{fig_simulated_nonthermal} (left panel) we show the fitted \veff\ profile
for the B09 simulation, along with its true distribution.
Since the velocity field predicted by the simulation was known, we were able to 
compute the true \veff\ profile by explicitly evaluating 
Eqn~\ref{eqn_veff} \citep[for more details of how this was done, see][]{fang09a}.
The fitted model was, by definition, a smooth function that cannot exactly capture
all of the features in the true profile, but the overall shape was well recovered,
especially in the vicinity of each spline knot. The largest deviations actually
occurred within the central radial bin, at scales that are, effectively, unresolved.
In the inner regions of the simulation (\ltsim 1~kpc) the 
X-ray isophotes  were quite flattened and there was a strong 
azimuthal temperature gradient (B09), so the spherical, single-phase approximation 
that is implicit in the deprojection method we adopted is a simplification. 
Despite this, the impact of these effects was modest and the recovered
\veff\ profile was sufficiently accurate for our purposes.

In Fig~\ref{fig_simulated_nonthermal} (right panel), we show the fitted 
\fnonthermal\ profile for the B09 simulation, and its true distribution. The 
nonthermal support in the simulation actually arises (mostly) from bulk 
gas motions rather than an additional pressure component. Still, \fnonthermal\
and \veff\ are highly degenerate, so we can still parameterize the 
nonthermal support as an ``effective'' \fnonthermal. Writing out 
Eqn~\ref{eqn_he2} for the two cases (\veff$\equiv 0$ and \fnonthermal$\equiv 0$), and 
eliminating $d P_g/dr$ between them, we find that \fnonthermal\ and \veff\
are related by:
\begin{equation}
v_{eff}^2 = f_{nth} v_c^2 +\frac{kT}{\mu m_H} \frac{d \log (1-f_{nth})}{d \log r} 
\end{equation}
We integrated this equation numerically to obtain the ``true'' \fnonthermal, which we found
to be in good agreement with the fitted profile 
(Fig~\ref{fig_simulated_nonthermal}).

%In practice, to 
%estimate \fnonthermal\ from the simulation, we actually calculated the difference
%between the spherically-averaged $P_g$ and the total pressure, 
%$P\equiv P_g/(1-f_{nth})$. 
%We used $P\simeq P_g(\infty)+\int_r^{\infty} dr v_c^2 <\rho_g>/r$, which comes from radially 
%integrating Eqn~\ref{eqn_vc} if ${\bf v}$ is assumed to be zero, and assuming \fnonthermal$=0$
%at large radii. This is, however, only approximate as the weighting used in the 
%spherical averaging of $P_g$ and $P$ are subtly different.
%the central bin. The largest discrepancy occurs between the spline knots, around
%$\sim$0.1~kpc, because the fitted model does not actually have sufficient 
%flexibility to produce this shape. Since this occurs at scales too small to 
%resolve, we decided against adding an additional
%spline knot at this scale to help capture this behaviour.  Overall, we conclude
%that the agreement between the true and fitted profiles is 
%sufficient for our purposes.

\section{Application to \src}
\subsection{Results}
\begin{figure}
\centering
\includegraphics[width=3.5in]{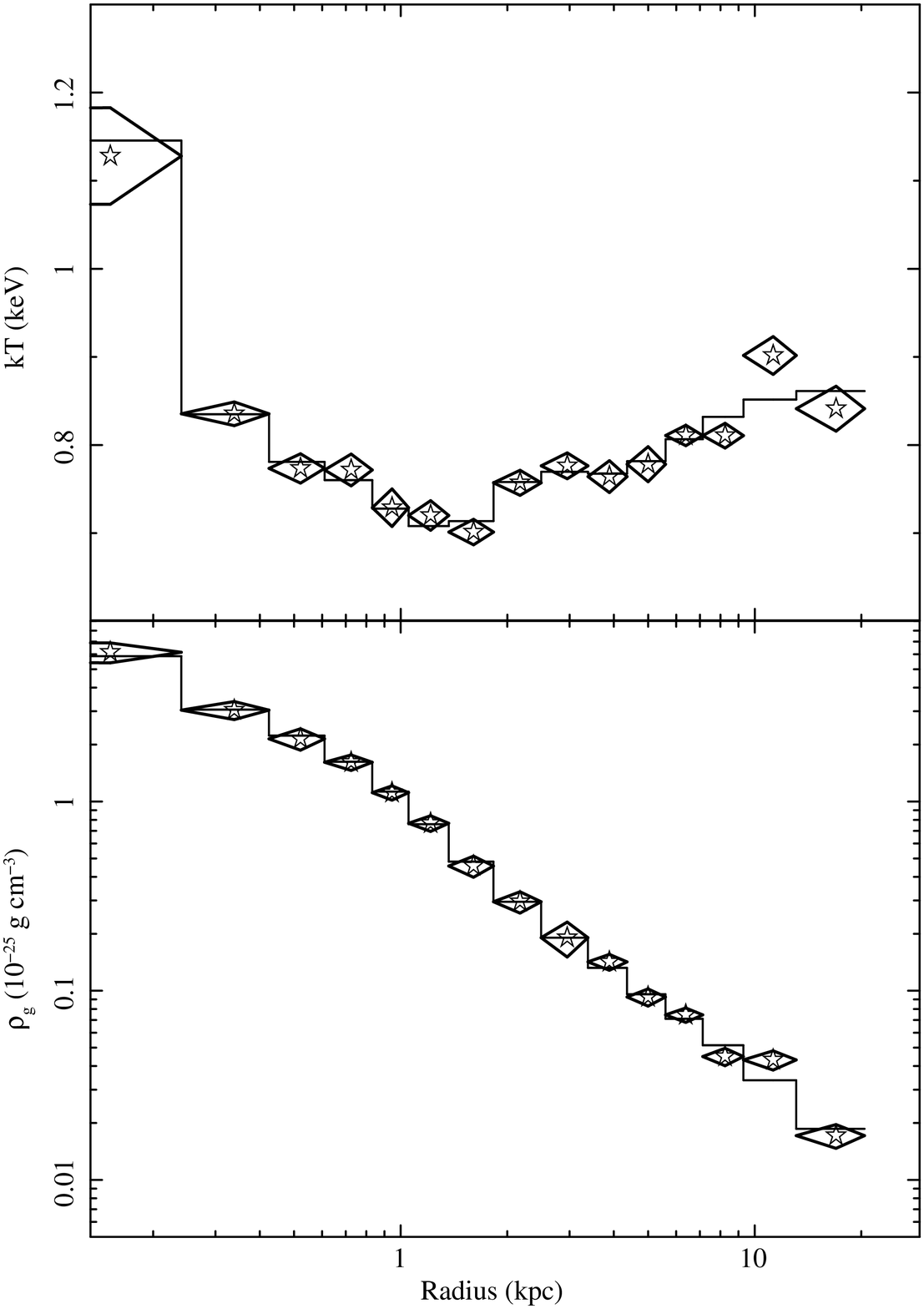}
\caption{Profiles of kT (upper panel) and gas density (lower panel), along
with the best-fitting model. Note the 
good agreement in all radial bins.\label{fig_kt_rho}}
\end{figure}
\begin{figure*}
\centering
\includegraphics[width=7in]{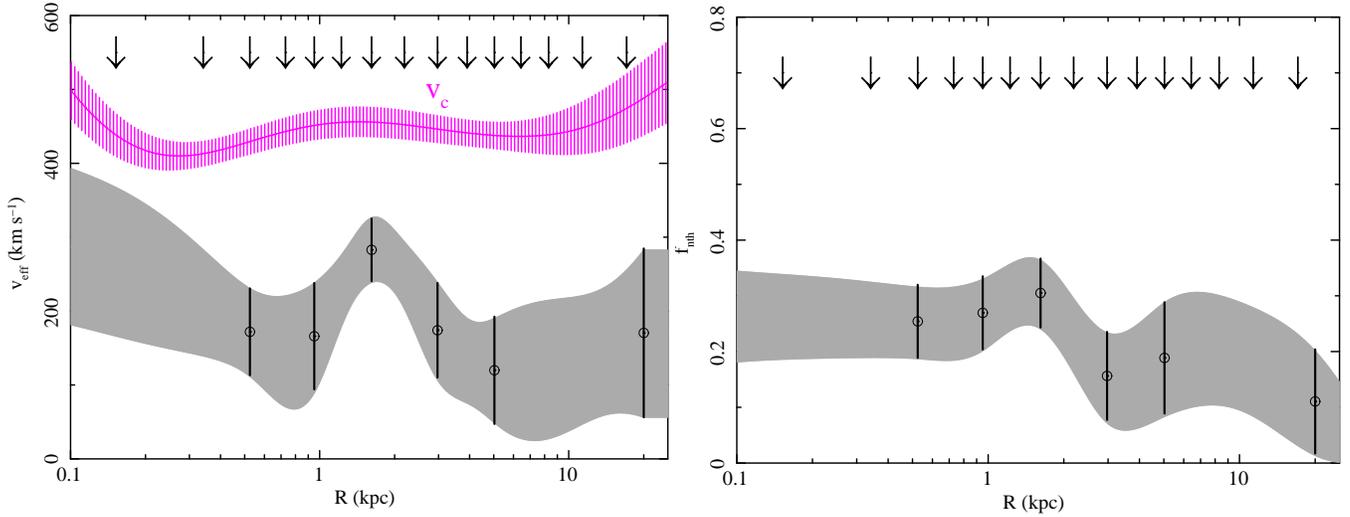}
\caption{The same as Fig~\ref{fig_simulated_nonthermal} but for the real \src\ data. 
In this case the ``true'' profile is not shown as it is {\em a priori} unknown,
and $v_c$ is derived from the joint fit to the stellar dynamical and X-ray data.
\label{fig_n4649_nonthermal}}
\end{figure*}
\begin{figure*}
\centering
\includegraphics[width=7in]{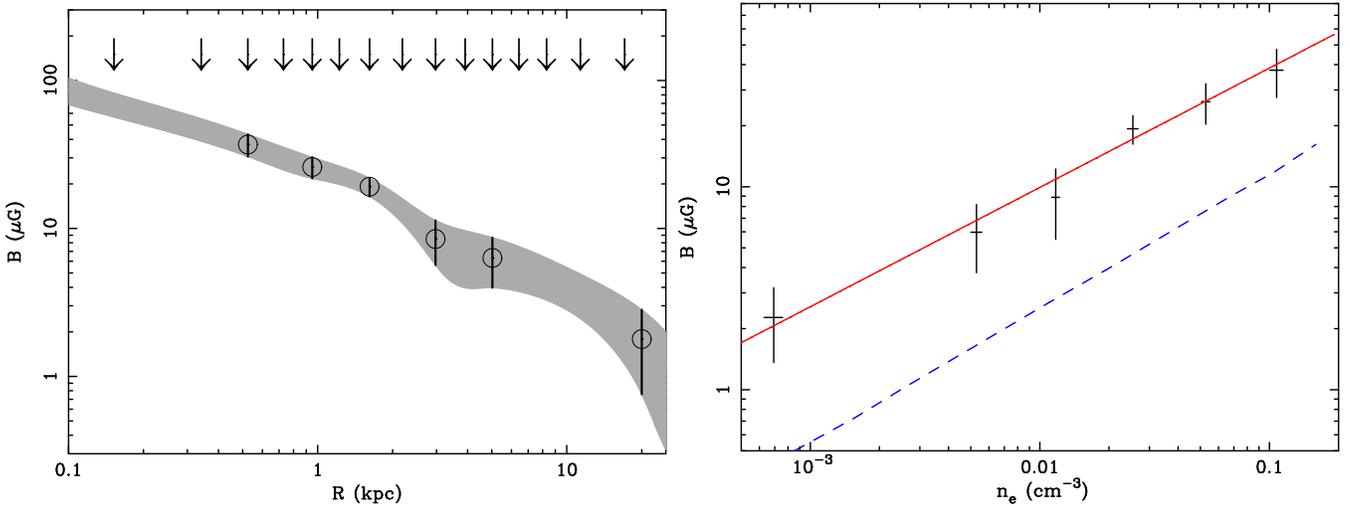}
\caption{{\em Left panel:} 
The implied magnetic field profile in \src, {assuming the nonthermal pressure
comes from the B-field in equipartition with $\sim$230~\kms\ random turbulence}.
The data-points are the fitted spline 
points, while the grey region is the 1-$\sigma$ confidence range, interpolating
between them. The arrows indicate the centres of each temperature and density 
data-bin. {\em Right panel:} The corresponding B-field {\em versus} electron density.
The solid red line is the best-fitting powerlaw relation (see text). The dashed
blue line is the prediction for the model of \citet[][their Fig~2]{mathews97a}. 
The model slope
agrees well with the observed data. {The quoted uncertainties do not include
systematic errors, including those from the stellar dynamical analysis. 
In particular, we note that, if the dynamical analysis of 
\citet{das11a} is taken at face value, the implied level of nonthermal support 
may be lower than inferred here (see text).}
\label{fig_n4649_magnetic}}
\end{figure*}
To study \src, we adopted the same
parameterized gravitating mass model as \citet{shen10a} (black hole, stellar component
and a cored logarithmic dark matter mass model), and folded in the results of the 
stellar dynamical analysis as priors on the black hole mass, stellar M/L ratio
and parameters of the dark matter model. For the stellar component, 
\citeauthor{shen10a} deprojected the V-band stellar light distribution from 
\citet{kormendy09a}, assuming an edge-on, oblate spheroidal geometry and a constant 
axis ratio of 0.9. Unlike the axisymmetric assumption of the dynamical analysis,
the X-ray technique assumes spherical geometry, and so it is important to 
average this profile spherically \citep{buote11b,buote11c}, which we did numerically.

We separately parameterized \veff\ and \fnonthermal\ as a 7-knot spline 
(each knot being placed approximately logarithmically in radius), and assuming
flat priors on the model values at each knot point, within reasonable limits. 
The best-fitting model actually captured the shapes of the density and temperature
profiles acceptably, {although the \fnonthermal\ parameterization is preferred}
($\chi^2$=18.7/13 and 22.1/13 for the \fnonthermal\ and \veff\
parameterizations, respectively; see
Fig~\ref{fig_kt_rho}). The resulting \veff\ and \fnonthermal\ 
profiles are shown in Fig~\ref{fig_n4649_nonthermal}.

The shape of the \veff\ profile is relatively flat, and could be consistent with
a constant gas rotation velocity (\veff=$232\pm37$\kms), or with a mildly falling 
profile
that becomes consistent with zero outside $\sim$5~kpc. Expressed in terms of 
\fnonthermal, the profile is also consistent both with a constant nonthermal 
pressure fraction (\fnonthermal$=0.27\pm0.06$) or with a gradually declining 
distribution that becomes consistent with zero outside $\sim$2~kpc. 
Still, within $\sim$2~kpc, approximately $\sim$30\%\ of the support must come 
from bulk gas motions or nonthermal pressure.

\subsection{Systematic error budget} \label{sect_syserr}
\begin{table*}
%{llllrrr}
%\tablecaption{\src: Results and error budget\label{table_results}}
%\tabletypesize{\scriptsize}
%\centering
%\rotate
\begin{tabular}{llllrrr}\hline
{Test} & {\veff(0.5~kpc)} & {\veff(1.6~kpc)} & {\veff(20~kpc)}& {\fnonthermal(0.5~kpc)} & {\fnonthermal(1.6~kpc)} & {\fnonthermal (20~kpc)} \\ 
& {(\kms)}  & {(\kms)}  & {(\kms)}  &  & & \\ \hline
Marginalized & $191 \pm 57.3$& $269^{+56.3}_{-28.9}$& $125^{+77.8}_{-116}$& $0.28^{+0.06}_{-0.08}$& $0.31 \pm 0.06$& $0.00^{+0.15}_{-0.000}$\\
 Best-fit& $(230)$& $(259)$& $(45.9)$& $(0.26)$& $(0.31)$& $(0.00)$\\\hline
$\Delta$Background&  $+25.8$  $\left( \pm {63}\right)$  &  $-11.4$  $\left( \pm {55.3}\right)$  &  $-52.4$  $\left( ^{+173}_{-73}\right)$ &  $-0.025$  $\left( \pm {0.06}\right)$  &  $-0.077$  $\left( \pm {0.06}\right)$  &  $\pm 0$  $\left( ^{+0.18}_{-0.00}\right)$ \\
$\Delta$Spectral&  $-34.6$  $\left( ^{+90.6}_{-24.9}\right)$  &  $+18.6$  $\left( ^{+40.2}_{-21.2}\right)$  &  $-125$  $\left( ^{+180}_{-0}\right)$ &  $+0.03$  $\left( ^{+0.05}_{-0.09}\right)$  &  $+0.02$  $\left( \pm {0.07}\right)$  &  $+0.003$  $\left( ^{+0.13}_{-0.00}\right)$ \\
%$\Delta$Distance &  $^{+20.2}_{-141}$  $\left( ^{+120}_{-53}\right)$  &  $-61$  $\left( ^{+50.8}_{-76.9}\right)$  &  $-125$  $\left( ^{+294}_{-6.34}\right)$ &  $+0.03$  $\left( ^{+0.05}_{-0.08}\right)$  &  $^{+0.00}_{-0.10}$  $\left( \pm {0.07}\right)$  &  $+0.10$  $\left( ^{+0.29}_{-0.10}\right)$ \\
$\Delta$Distance &  $^{+20.2}_{-67.9}$  $\left( ^{+53.8}_{-107}\right)$  &  $-97.9$  $\left( \pm {54.7}\right)$  &  $-125$  $\left( ^{+294}_{-23.7}\right)$ &  $+0.03$  $\left( ^{+0.05}_{-0.08}\right)$  &  $^{+0.00}_{-0.10}$  $\left( \pm {0.07}\right)$  &  $+0.10$  $\left( ^{+0.29}_{-0.10}\right)$ \\
$\Delta$Knots&  $+58.2$  $\left( ^{+37.9}_{-53.5}\right)$  &  $^{+11.7}_{-20.2}$  $\left( ^{+41}_{-53.5}\right)$  &  $\pm 120$  $\left( ^{+114}_{-53.5}\right)$ &  $+0.02$  $\left( ^{+0.05}_{-0.09}\right)$  &  $-0.043$  $\left( \pm {0.07}\right)$  &  $+0.28$  $\left( \pm {0.07}\right)$ \\
$\Delta$Fit priors&  $^{+33.3}_{-24.5}$  $\left( \pm {93.1}\right)$  &  $^{+126}_{-12.6}$  $\left( ^{+47.5}_{-112}\right)$  &  $^{+188}_{-86.3}$  $\left( \pm {169}\right)$ &  $^{+0.03}_{-0.02}$  $\left( ^{+0.08}_{-0.11}\right)$  &  $^{+0.05}_{-0.01}$  $\left( \pm {0.10}\right)$  &  $+0.06$  $\left( ^{+0.15}_{-0.06}\right)$ \\
$\Delta$Entropy &  $+1.69$  $\left( \pm {60.6}\right)$  &  $-10.8$  $\left( ^{+58.2}_{-26.4}\right)$  &  $-67$  $\left( ^{+125}_{-58.3}\right)$ &  $+0.01$  $\left( ^{+0.04}_{-0.10}\right)$  &  $+0.03$  $\left( \pm {0.09}\right)$  &  $+0.006$  $\left( ^{+0.16}_{-0.01}\right)$ \\
$\Delta$Covariance &  $^{+29}_{-35.1}$  $\left( ^{+99.9}_{-59.3}\right)$  &  $\pm 19.2$  $\left( ^{+77.1}_{-54.5}\right)$  &  $-122$  $\left( ^{+125}_{-70.8}\right)$ &  $^{+0.03}_{-0.00}$  $\left( ^{+0.09}_{-0.12}\right)$  &  $+0.05$  $\left( ^{+0.10}_{-0.07}\right)$  &  $+0.001$  $\left( ^{+0.05}_{-0.00}\right)$ \\
$\Delta$Weighting&  $-9.86$  $\left( \pm {49.4}\right)$  &  $+47.8$  $\left( ^{+36.8}_{-53.9}\right)$  &  $-63.7$  $\left( ^{+377}_{-61.7}\right)$ &  $-0.013$  $\left( ^{+0.06}_{-0.07}\right)$  &  $+0.02$  $\left( ^{+0.06}_{-0.07}\right)$  &  $\pm 0$  $\left( ^{+0.14}_{-0.00}\right)$ \\ \hline
\end{tabular}
\caption{{Marginalized values and 1-$\sigma$ confidence regions for \veff\
and \fnonthermal\ measured at three representative radii. 
Since the best-fitting parameters need not be identical to the marginalized
values, we also list the best-fitting values for each parameter (in parentheses). In addition to the statistical errors, we also show estimates of the error budget
from possible sources of systematic uncertainty.
We consider a range of different
systematic effects, which are described in detail in \S~\ref{sect_syserr}; 
specifically we evaluate the effect of treatment of the background 
($\Delta$Background), spectral-fitting choices ($\Delta$Spectral), 
distance uncertainties ($\Delta$Distance), the number of spline knots
used in the calculation ($\Delta$Knots), priors on the fit parameters
($\Delta$Fit priors), the entropy parameterization ($\Delta$Entropy),
treatment of the covariance between data-points ($\Delta$Covariance)
and disabling the emissivity computation ($\Delta$Weighting).
We list the change
in the marginalized value of each parameter for every test and, in parentheses,
the statistical uncertainty on the parameter determined from the test.
Note that the systematic error estimates should {\em not}
in general be added in quadrature with the statistical error}\label{table_results}}
\end{table*}
{All methods for inferring the mass distributions in early-type 
galaxies entail arbitrary analysis choices, some of which may
quantitatively affect the measured mass. In this section,
we briefly explore how sensitive ourn conclusions are to the 
choices in the X-ray modelling. Further systematic uncertainties
may be associated with choices in the stellar dynamical analysis
(for example, the inclination angle assumed), but exploring them
is beyond the scope of the present paper.
The systematic error budget is summarized in Table~\ref{table_results},
and we discuss below how each test was performed.
For a more detailed discussion of the various systematic
error assessments considered here, see \citet{humphrey12a}.

We first examined the sensitivity of our spectral-fitting 
results to the treatment of the \chandra\ background 
(``$\Delta$Background'') by 
using the standard ``background template'' spectra, suitably
renormalized to match the data at \gtsim 10~keV, instead of the 
more robust, modelled background adopted by default
\citep{humphrey08a}. We experimented with replacing the 
APEC thermal plasma model used in spectral fitting with the 
MEKAL model (``$\Delta$Spectral'') and varying the distance
by $\sim$33\%\ \citep[corresponding to the statistical
error on the distance to \src\ given by][``$\Delta$Distance'']{tonry01}.
To investigate how changing the parameterized model used to 
fit the \veff\ or \fnonthermal\ profiles can affect the results,
we experimented with reducing the number of 
spline knots to 3 (at 0.5~kpc, 0.16~kpc and 20~kpc), 
and also with using a constant \veff\ or \fnonthermal\
(``$\Delta$Knots''). These simpler parameterizations did not lead to 
significantly poorer fits. 
Next we examined how the choice of priors might be influencing
our results by replacing each of the flat priors we adopted
in our analysis by priors which were flat in logarithmic space
(``$\Delta$Priors''). We also considered changing the entropy
parameterization by adding an additional break at large 
radii (``$\Delta$Entropy''), experimented with folding in the
full covariance between adjacent temperature and density data-points
(``$\Delta$Covariance'') and turning off the emissivity computation
when evaluating the weighted temperature and density models
(``$\Delta$Weighting''; see
\citealt{humphrey12a} for more details).
We list the largest changes and statistical errors associated
with each test in Table~\ref{table_results}. 
In almost all cases, the change in the 
marginalized value was not much larger than
the statistical error; in a few cases (especially at $\sim$20~kpc)
the statistical error-bars were enlarged by the test. Ultimately,
none of the statistical uncertainties were large enough to lead to
qualitatively different conclusions. Nonthermal support,
approximately at the $\sim$25\%\ level, is needed to reconcile the 
X-ray data with the published dynamical mass of \src.}

\section{Discussion}
\subsection{A non-hydrostatic X-ray model}
We have demonstrated that a simple modification to the entropy-based hydrostatic,
forward fitting technique 
\citep{humphrey08a,humphrey09d} allows nonthermal
pressure profiles to be inferred from X-ray data, provided additional,
unbiased, constraints on the mass profile are available (\eg\ from 
stellar dynamical mass measurements). At scales resolved by the density 
and temperature profiles, the new technique was able to recover 
reasonably well the radial distributions of \fnonthermal\ or \veff\ 
in a simulated galaxy, which was tailored to match real observations 
of the representative, bright system \src. At unresolved scales, 
the recovered distributions of 
\veff\ and \fnonthermal\ were approximately correct, but were sensitive to the 
exact parameterizations used to fit them.

The approach outlined here did not allow us to break the degeneracy between
bulk gas motions (\veff) and nonthermal pressure (\fnonthermal), and so 
we advocated determining each profile separately. In a real system, it is likely
that some combination of both effects could contribute to nonthermal support,
and so it is desirable to fold in additional constraints to try to disentangle
them. This can be easily implemented in a Bayesian framework by modifying the 
priors appropriately. In principle, resonance scattering measurements,
or spectral line broadening constraints, could be employed in this way as 
restrictions on the turbulent pressure component of \fnonthermal. 
Similarly, magnetic field rotation measure constraints could be employed
to restrict the magnetic pressure contribution to \fnonthermal.
As discussed in B09, the ellipticity profile of the gas actually provides
important information on the gas motion, potentially allowing it to be 
used as an indirect constraint on \veff. To do this fully self-consistently, as in 
B09, involves running a suite of hydrodynamical models, 
making it impractical to cover a suitably large region of parameter space.
With significant simplifications, for example assuming that the only gas motions are
rotational, the problem becomes tractible. Unfortunately, to evaluate the 
models properly then involves relaxing the spherical approximation, which
substantially complicates the analysis beyond the scope of this
paper. Non-spherical hydrostatic models
including rotation have been constructed before \citep[\eg][]{buote96a,statler02a},
but not for computing detailed temperature and density distributions.

As presented here, the model provides a means for measuring deviations
from hydrostatic equilibrium in the gas. Conversely, if a physical model
exists for \veff\ or \fnonthermal, the same approach could be used to correct
the X-ray mass measurement for deviations from hydrostatic equilibrium,
making X-ray studies alone of nearby galaxies potentially suitable for high-precision 
cosmology. At the moment, our understanding of nonthermal support in the centres of 
galaxies is still in its infancy, but when the \veff\ and \fnonthermal\ profiles of 
sufficiently large a sample of systems have been measured, this may become routine.
\subsection{Nonthermal support in \src}
Applying our model to  \src, 
we have carried out a joint stellar dynamical and X-ray mass analysis,
effectively reconciling the measurements made by \citet{shen10a} and 
\citet{humphrey08a} with a modest but significant 
($\sim$25\%) nonthermal pressure profile.
{We obtained these estimates by assuming that the ISM is maintained in static equilibrium, whereas in practice, we expect there to be a modest cooling flow (\eg\ \citetalias{brighenti09a}). 
The measured level of nonthermal support will then be an overestimate, albeit only very slightly if the cooling flow is very subsonic}.
By comparing the mass profiles obtained from stellar dynamical and X-ray analysis 
(Fig~\ref{fig_vc}), \citet{das10a} similarly concluded that the nonthermal pressure
in NGC\thin 4649 is no larger than $\sim$32\%, although they did not present a detailed
\fnonthermal\ profile.
{\em This level of deviation from hydrostatic equilibrium is fully 
comparable to that expected in galaxy clusters, which are routinely used for cosmology}
\citep[\eg][]{tsai94a,buote95a,evrard96a,rasia06a,nagai07a,piffaretti08a,fang09a,lau11a}, and is 
consistent with (albeit slightly larger than inferred in) the conclusions of 
\citet{churazov08a} for two more morphologically disturbed systems.
Considering the level of statistical and systematic errors in our previous
studies of early-type galaxies \citep[\eg][]{humphrey06a,humphrey08a,humphrey09d,humphrey11a,humphrey12a,humphrey12b}, this level of nonthermal pressure, if ubiquitous, 
would not imply qualitatively different conclusions on the gravitating mass profiles.
%DONE

{Since the level of nonthermal support is inferred from the subtle differences
between the mass profiles derived with two techniques, control of the 
systematic errors is important for a robust measurement. Certain choices,
such as the treatment of the X-ray background, the adopted fit priors, and 
the number of spline knots used, can have a non-negligible impact on the 
recovered profile, as is clear from \S~\ref{sect_syserr}. Given present
uncertainties (both statistical and systematic)}, the data currently do not 
allow us to determine whether the nonthermal support
profiles (\veff\ or \fnonthermal) are constant with radius 
\citep[as assumed by][]{churazov08a}, or if they are gradually declining; 
they are clearly consistent with zero outside $\sim$5~kpc. While we 
anticipate improved {statistical} X-ray constraints if we fold in the available archival \xmm\ data and the significantly deeper \chandra\
observations ($\sim$200~ks, as compared to the 81~ks used in our 
analysis) that will soon be publicly available in the archive, {the 
large statistical uncertainty on the current optically inferred mass 
at large radii (Fig~\ref{fig_vc}) is a major limiting factor.}

At present, we have not folded in any systematic errors in the stellar dynamical
analysis into our study. In particular, uncertainties in the inclination of the 
system may be important \citep[\eg][]{thomas07b}. \citet{shen10a} explored
different inclination angles with a simplified model that included a fixed dark matter
halo, finding that inclination had little impact on their results. 
Given the restricted parameter space explored in this
way, we did not attempt to incorporate these fits onto our work 
and marginalize over inclination. 
{More intriguing are the recent particle based models of \citet{das11a}, who
found much larger confidence regions for the enclosed mass profiles than 
\citeauthor{shen10a} when 
fitting kinematic data for the stars and planetary nebulae, implying a 
lower level of nonthermal support than measured here. Some of the 
discrepancies with the \citeauthor{shen10a} results may arise in part 
due to there being different equilibrium configurations for the globular
cluster and the planetary nebulae populations. However, this should not
be important in the central part of the galaxy, where the inferred
nonthermal pressure is strongest. This is supported by the good agreement 
between the stellar M/L ratios inferred by \citeauthor{shen10a} and 
\citet{gebhardt03a}, who used only stellar kinematics.
While we expect the state of the art dynamical modelling of
\citeauthor{shen10a} to be robust, it is unclear
why the code used by \citeauthor{das11a} produced such a different result. 
Until this tension is resolved, some questions persist over the exact
level of nonthermal support in \src.} {As it stands, the true level of 
nonthermal support may be lower than we infer}. 
Nevertheless, in the following sections, 
we consider the 
physical implications of nonthermal support at the level inferred in
our study.

%{\bf In what follows, we assume that nonthermal support is 
%sufficient to maintain the ISM in a static state, whereas we expect a modest
%inflow \citep[\eg][]{brighenti09a}.}

\subsubsection{Gas rotation}
Given the smooth, relaxed isophotes of \src, we would not expect large-scale bulk gas 
motions other than, perhaps, gas rotation, which is expected in the centre
of a cooling flow \citep[\eg][]{nulsen84a,kley95a,brighenti96a}. 
B09 computed detailed hydrodynamical
models for the gas flow in \src, which they tailored to match the gas density,
temperature and ellipticity profiles. They 
concluded that rotation could be dynamically
important within the central $\sim$kpc. Comparison of their inferred 
\veff\ profile (Fig~\ref{fig_simulated_nonthermal}, left panel) to the 
real data (Fig~\ref{fig_n4649_nonthermal}, left panel) strongly suggests that,
while rotation could dominate the nonthermal support at these small scales,
it cannot explain the observations at larger scales ($\sim$1--3~kpc).
It is likely that B09 underestimated the level of gas rotation, as they 
adopted the gravitational potential inferred from the hydrostatic fit,
rather than the deeper (true?) potential from dynamics, but the modest
underestimate in \veff\ that is implied {in order to maintain the same level of isophotal flattening} (\ltsim 20\%) will still not 
reconcile the simulated rotation velocity with the observed profile
at these scales.

\subsubsection{Random turbulence}
Alternatively, the nonthermal pressure profile (\fnonthermal) could 
arise due to turbulent motions within the gas. Assuming isotropic
turbulence and eddy velocities drawn from a Maxwell-Boltzmann
distribution, the turbulent pressure is given by 
$P_{t} \simeq (1/3) \rho_g v_{turb}^2$, 
where $v_{turb}$ is the r.m.s.\ turbulent velocity. Adopting 
kT$=$0.8~keV, we find that $v_{turb} \simeq 600 \sqrt{f_{nth}/(1-f_{nth})}$~\kms,
or $\sim$360$\pm$60~\kms\ for \fnonthermal=$0.27\pm0.06$. 
{Random turbulence of this magnitude may help to 
generate a dynamically important magnetic field that would also
contribute nonthermal pressure (see \S~\ref{sect_bfields}).
Whether plausible sources of random turbulence, such 
as mixing from stellar mass loss and Type Ia supernovae, galaxy
merging and ``sloshing'', or AGN-driven disturbances in the ISM
are sufficient to maintain turbulence at this level is unclear.

Turbulent diffusion can provide a means for transporting angular momentum outwards, counteracting the ISM rotation that is expected to be induced by inflow, and thus making the X-ray image rounder \citep{brighenti00a}. 
Although sensitive to the assumed length-scale of the largest turbulent eddies, 
\citetalias{brighenti09a} used the isophotal flattening in the core of \src\ to
infer turbulent velocities $\sim$50~\kms, which are not dynamically important.}

It is worthwhile to compare the inferred level of turbulence with that 
implied in the centres of elliptical galaxies from other means. 
\citet{sanders11a} placed an upper limit of 700\kms\ on
line broadening in the \xmm\ RGS spectrum of \src, which is too large
to be interesting.
While \citet{werner09a} were not able to place useful constraints
on resonance scattering in \src, they measured the effect in three other
similar galaxies, including NGC\thin 4636, for which they inferred $v_{turb}<100$\kms. 
Conversely \citet{deplaa12a} inferred significant turbulence ($\sim$140--700\kms)
in two other systems, which may be consistent with our observation
of \src. However, we note that, 
for one of these systems, NGC\thin 5044, \citet{sanders11a} concluded 
that the width of the spectral lines in RGS spectra was fully consistent 
with the source just being spatially extended, leaving little room for 
turbulent broadening. Whether turbulence as large as $\sim$360\kms\ 
is realistic in the centre of a giant elliptical galaxy therefore remains 
unclear at this point, but it does not appear ubiquitous. Still, this may
be resolved by observations with \astroh\ (\S~\ref{sect_astroh}).

\subsubsection{Cosmic Ray Pressure}
Another possible cause of nonzero \fnonthermal\ is cosmic ray pressure. Strong
cosmic ray injection into the ISM from the radio jet is likely to 
inflate large cavities \citep[\eg][]{mathews08b}, which are not seen
in \src\ \citep{humphrey08a}. Still, it remains plausible that 
there could be non-negligible cosmic ray pressure in the vicinity of the 
(weak) jets, which extend for \ltsim 1.5~kpc \citep{shurkin07a}. 
Intriguingly, our current measurement (Fig~\ref{fig_n4649_nonthermal}) is 
consistent with \fnonthermal\ being significantly weaker outside 
the central $\sim$2~kpc. In fact, the fit is formally indistinguishable
if the nonthermal pressure is set to zero at these scales 
($\chi^2$/dof=19.6/16, as compared to 18.7/13).
%evidence is -34.7, and -38.0, respectively 
If cosmic ray injection is primarily responsible for the nonthermal
support in \src\ and similar objects, we would expect to see a 
strong correlation between the radii requiring \fnonthermal$\ne0$ and the 
extent of the jets. This is potentially testable as the jet morphology
differs widely in nearby galaxies that are accessible for this kind of 
joint X-ray-optical analysis.

\subsubsection{Magnetic Fields} \label{sect_bfields}
{Observationally, there are known to be \gtsim $\mu$G magnetic fields 
in the centres of some early-type galaxies \citep{vallee11a},
which may provide an additional source of nonthermal pressure.
For an early-type galaxy, a plausible mechanism for maintaining 
such a field would be the turbulent dynamo effect
\citep[\eg][]{mathews97a,brandenburg05a}. Assuming that the 
entire nonthermal pressure is provided by this mechanism, and assuming
equipartition between the magnetic and turbulent energy
densities (which should maintain a disordered field), 
we computed the B-field required in \src, using
$P_{mag}=B^2/8\pi$, where $P_{mag}$ is the magnetic pressure and $B$ is the 
B-field in cgs units.
The implied field strength profile, shown in 
Fig~\ref{fig_n4649_magnetic} (left panel), falls
steeply with radius (roughly $B=24 (R/kpc)^{-0.8} \mu G$) and requires
turbulent velocities, $v_{turb}\simeq 230$~\kms.  The radial
B-field dependence is similar to what is expected from simple galaxy formation
models \citep{mathews97a,beck12a}, and comparable fields have been inferred
at small scales in a few galaxies with embedded radio jets
\citep{vallee11a}, albeit such strong fields are not ubiquitous.

In the right hand panel of Fig~\ref{fig_n4649_magnetic}, 
we show how the implied B-field varies 
with the electron density. The data were well fitted 
($\chi^2$/dof=1.1/4)
%\footnote{We ignore the data-point corresponding to the 
%highest densities in this fit, as it is inferred by extrapolating the fitted
%model to the smallest, unresolved, scales.} 
with a model 
$B=B_0 \left( n_e / 0.1 cm^{-3} \right)^\gamma$, where 
$B_0=39\pm6 \mu$G and $\gamma=0.59\pm0.09$. 
We note that this is similar to the predictions of magnetic field
generation in a (nonrotating) cooling flow predicted by \citet{mathews97a}, albeit
the field strength is much higher (Fig~\ref{fig_n4649_magnetic}). 
Faraday rotation measure constraints in some galaxy clusters imply 
$\gamma$ ranging from $\sim$0.5--1.0 \citep[\eg][]{dolag01a,guidetti08a},
similar to our inferred value for \src.
}
% Note: fits powerlaw of form B=B0(ne/0.1)^alpha gives:
% B0=49.665       -7.6266        8.1264  mu G (1-sigma)
% index = 0.58809 -0.79823E-01   0.90083E-01

\subsection{Measuring gas motions with \astroh} \label{sect_astroh}
\begin{figure}
\centering
\includegraphics[height=3.5in,angle=270]{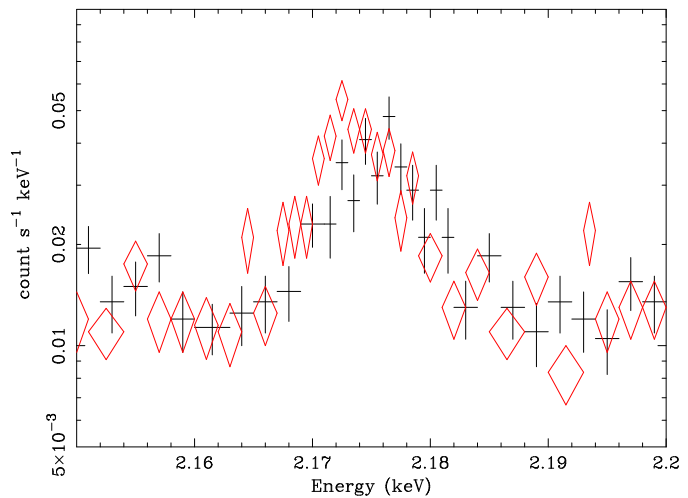}
\caption{Portions of two simulated \astroh\ SXS spectra, showing a 
Si XIII line. The spectra were accumulated from the entire SXS field 
of view in two different pointings, which were centred at different 
positions along the major axis of \src, symmetrically about the core
of the galaxy.  The simulations assumed that nonthermal support was entirely
due to gas rotation; the Doppler shifting of the X-ray lines
produces a clear shift in the line centroid between the 
two spectra (shown as crosses and diamonds). For illustration purposes, 
the simulated spectra shown have 1~Ms exposures, as opposed to the 200~ks
exposures discussed in the text.\label{fig_astroh}}
\end{figure}
If turbulence or bulk gas motions are primarily responsible for
the nonthermal support in \src, we should 
see associated Doppler broadening or centroid shifts of the X-ray
emission lines. Given the spatially extended nature of the hot gas,
no currently in-orbit instrument would allow these modest shifts to be
measured definitively. However, the nondispersive, high resolution spectroscopy 
enabled by microcalorimeters scheduled to fly on future
missions is ideally suited to this task. 

Of particular interest is the Soft X-ray Spectrometer (SXS) aboard 
the \astroh\ observatory \citep{takahashi10a}. To explore whether 
it will be possible to measure the predicted gas motions in \src\ with
this instrument, we simulated two deep ($\sim$200~ks)
exposures, each centred along the major axis at a distance $\sim$1.5\arcmin\ 
from the galaxy core. Spectra were extracted from the entire 3$\times$3\arcmin\
field of view in each case. If the gas is rotating, one spectrum 
should be redshifted, and one blueshifted.
To simulate the spectra, we modified the 
Monte Carlo approach outlined in \S~\ref{sect_simulations}. 
At a series of grid positions (R,$\theta$,z), we generated
photons, assuming a thermally broadened APEC plasma with a given 
density, temperature,abundance, line-of-sight velocity and 
(turbulent) Doppler line broadening. These
parameters were derived from the best-fitting models, assuming 
spherical symmetry. Assuming isotropy, turbulence should produce
Gaussian broadening of the line, 
$I\propto exp \left( -(E-E_0)^2/\Delta E^2\right)$, where 
$E_0$ is the line centre, and 
$\Delta E/E=\sqrt{2/3} v_{turb}/c \equiv b/c$, where b is the 
Doppler b-parameter.

To incorporate the sensitivity of the SXS, we 
used the standard ``baseline'' on-axis \astroh\ responses,
{\tt ah\_sxs\_7ev\_basefilt\_20090216.rmf} and {\tt sxt-s\_100208\_ts02um\_of\_intallpxl.arf}\footnote{\href{http://astro-h.isas.jaxa.jp/researchers/sim/response.html}{http://astro-h.isas.jaxa.jp/researchers/sim/response.html}}. Photons were 
then projected onto the sky, folding in the nominal on-axis 
point spread function (the {\tt ``ah\_sxt\_psfmodel\_20090217''} model)\footnote{\href{http://astro-h.isas.jaxa.jp/researchers/sim/GSFC_mirror.html}{http://astro-h.isas.jaxa.jp/researchers/sim/GSFC\_mirror.html}}.
Additional background photons were generated, corresponding to
emission from low mass X-ray binaries (a 7.3~keV bremsstrahlung
model distributed like the stellar light and normalized according to
the measurement of \citealt{humphrey08b}), the cosmic X-ray
background (a powerlaw with $\Gamma$=1.41 and normalization given by
\citealt{deluca04a}), the Galactic foreground 
(a 0.07~keV and a 0.20~keV APEC plasma component, \eg\ \citealt{humphrey11a})
and emission from the Virgo cluster ICM 
(a 2.5~keV APEC component; \citealt{humphrey08a}). For simplicity, we
ignored the (featureless) instrumental background which is at least 
an order of magnitude fainter than the source emission below $\sim$3~keV.

Since there was a range of gas temperatures projected into the
field of view, we modelled the background-subtracted spectrum using 
two APEC models with (the same) Doppler broadening and tied abundances
and redshifts, plus a bremsstrahlung model to account for LMXBs. We
found this model adequately fitted the spectra. If gas rotation supplies
the nonthermal support, we would expect a peak $\sim$500\kms\ velocity 
gradient between the two fields, which corresponds to $\sim$4~eV at
2.2~keV. In practice, the large apertures used caused the line profiles
to be smeared out. Nevertheless, as shown in Fig~\ref{fig_astroh},
the line centroid shift should be detectable; for 200~ks exposures, we 
expect to detect the gas rotation at $\sim$4-$\sigma$.
Conversely, if turbulence dominates the gas dynamics, we would expect
$\sim$200\kms\ broadening of the lines, and little velocity gradient.
Simulating corresponding spectra, we found the line broadening
was required; we constrained the line of sight rms velocity to
$180\pm40$\kms.
{\em Therefore, if gas motions are responsible for the nonthermal support in \src, we expect to be able to measure them with \astroh}.

\section*{Acknowledgments}
%\acknowledgements
We would like to thank Taotao Fang and Fabio Gastaldello for 
discussions.
%This research has made use of data obtained from the High Energy Astrophysics
%Science Archive Research Center (HEASARC), provided by NASA's Goddard Space
%Flight Center.
%This research has also made use of the
%NASA/IPAC Extragalactic Database (\ned)
%which is operated by the Jet Propulsion Laboratory, California Institute of
%Technology, under contract with NASA, and the HyperLEDA database
%(http://leda.univ-lyon1.fr). 
%We are grateful to the ACE/ SWICS(????) instrument
%team for making their data publicly available through the ACE Science Center.
PJH and DAB gratefully acknowledge partial support from NASA under Grant NNX10AD07G, 
issued through the office of Space Science Astrophysics Data Program. 
%Partial
%support for this work was also provided by NASA under Grant 
%NNG04GE76G issued through the Office of Space Sciences Long-Term
%Space Astrophysics Program, and by \chandra\ award ??? and NASA-\xmm\ grant
%???.

\appendix
\section{Comparison with X-ray analysis of Das et al.}
\begin{figure*}
\centering
\includegraphics[width=7in]{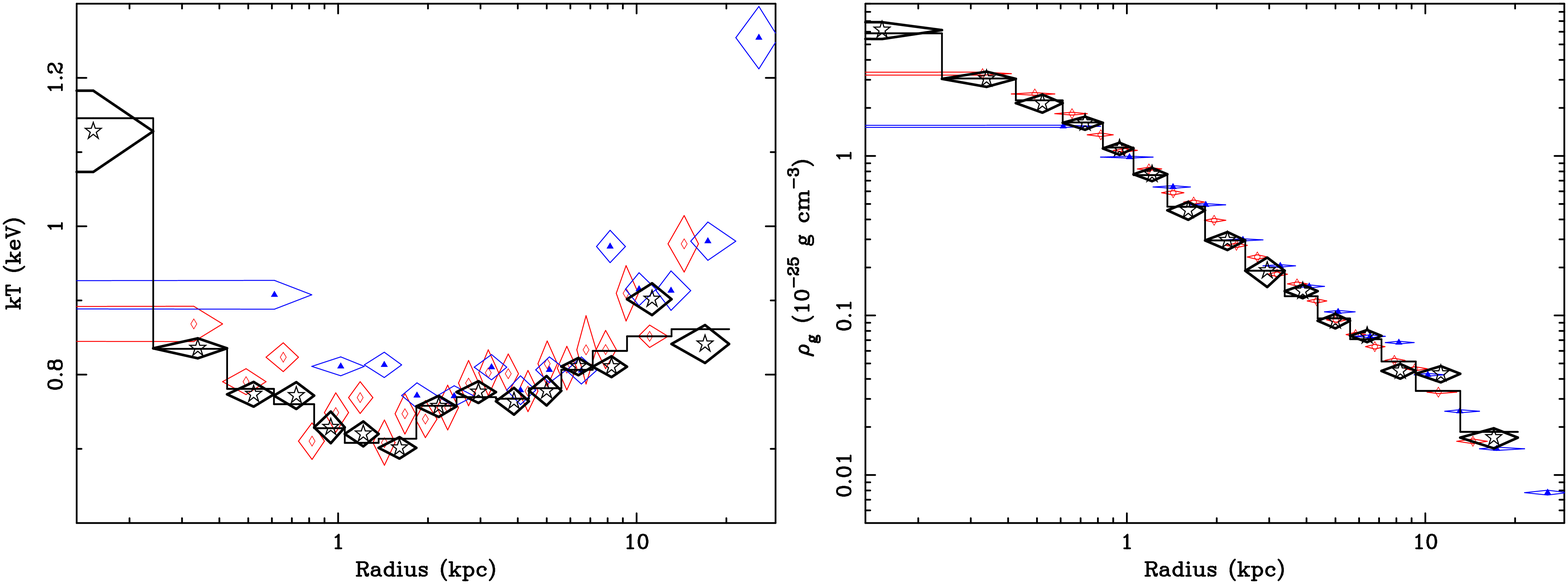}
\caption{{\em Left panel:} \chandra\ temperature profile obtained by 
\citet{humphrey08a} (black stars), compared to the profiles used 
by \citet{das10a} (\chandra: red diamonds, \xmm: blue triangles). Note
the general agreement between the \chandra\ profiles inside $\sim$10~kpc, 
but the disagreement with \xmm\ at the smallest scales. {\em Right panel:}
The same for the gas density profiles. Since the density profiles were 
shown with arbitrary normalization in \citet{das10a}, we have arbitrarily 
scaled them here for clarity.
\label{fig_compare_das}}
\end{figure*}
\citet{das10a} introduced a minimally parametric method for recovering the 
mass profiles of systems in which the gas is approximately hydrostatic,
by inverting the temperature and density profile data. Their approach is 
a variant of the ``smoothed inversion'' technique \citep[for a critical 
review of mass modelling techniques, see][]{buote11a}, 
in which, rather than 
fitting smooth models to the density and temperature profile, they 
instead {apply a  smoothing prior, controlled by the parameter $\lambda$,
which is calibrated against realistic data. 
They assumed that $\lambda$ 
is the same (and fixed) for all observables (temperature, density, rotation 
velocity) and all systems. The advantage of this method is that it makes few 
{\em a priori} assumptions about the shape of the mass or temperature profiles. 
The disadvantage is that it assumes that all of the profiles are smooth in
a particular manner, which is the same for all profiles\footnote{Compare this
with the forward fitting method, which imposes (generally smooth) functional forms
on the gravitating mass and entropy profile.}.}

In Fig~\ref{fig_vc}, we compare the $v_c$ profiles obtained by our method
and by \citet{das10a}. In general, the agreement is good over the range
$\sim$3--9~kpc, but the mass found by \citet{das10a} is higher outside of this
range. In Fig~\ref{fig_compare_das}, we compare the temperature and density
data which were fitted in our analysis, and those used by \citeauthor{das10a}. 
It is immediately
clear that there is good agreement between all three data-sets between
$\sim$3--9~kpc, where the $v_c$ profiles also agree, but discrepancies 
arise outside this range.
We conclude that the differences between the mass inferred from the hydrostatic
analysis of \citet{das10a} and \citet{humphrey08a} are most probably a 
consequence of the different {\em temperature and density profiles}, 
rather than intrinsic differences in the mass fitting techniques.

At the smallest scales, there is a large difference between the \xmm\ and 
\chandra\ data-points, which probably reflects the larger PSF of \xmm\
(90\%\ encircled energy radius$\simeq$30\arcsec$=$2.2~kpc) compromizing
the inner data by spectral mixing between adjacent annuli. This offset
may explain the formally poor reduced $\chi^2$ ($\simeq$6) obtained by 
\citeauthor{das10a} in their fits. We note that such large systematic 
errors will distort the $\chi^2$ topology, even if the best-fitting model
is correct, compromizing the error-bar calculation.

The remaining differences with our \chandra\ profiles (in particular the 
lower temperature we found at large scales) probably 
reflect different analysis choices made by \citet{churazov10a},
who derived the profiles used by \citeauthor{das10a}. Specifically,
in their analysis the ISM abundance was fixed to a constant value
(\zfe=0.5), and Solar abundance ratios were used for all species, which 
is not formally correct for \src\
(\citealt{humphrey05a,humphrey06a}; \citetalias{brighenti09a}). Furthermore, they ignored
the contribution of unresolved LMXBs (which can affect the inferred
density) and projected emission from the Virgo ICM (which can affect the 
temperature) in their fits, while their correction for projected
emission from gas beyond the outermost shells used in the deprojection
did not account for the truncation of the halo by the Virgo ICM. 
In our modelling,
we explicitly accounted for these effects \citep{humphrey08a}.
Nevertheless, even with the different analysis choices, we see that the 
recovered mass profile is not dramatically affected over most of the 
interesting radial range (Fig~\ref{fig_vc}).

%\clearpage

%\bibliographystyle{apj_hyper}
%\bibliographystyle{apj_nohyper}
\bibliographystyle{mnras_hyper}
\bibliography{paper_bibliography.bib}

\end{document}